\documentclass[9pt,twocolumn,twoside]{opticajnl}
\journal{opticajournal} 

\setboolean{shortarticle}{true}

\usepackage{lineno}
\usepackage{graphicx}
\usepackage{tabularx}
\usepackage[utf8]{inputenc}

\linenumbers 

\title{Efficient broadband antenna for a quantum emitter working at telecommunication wavelengths}

\author[1]{Monika Dziubelski}
\author[1]{Joanna M Zajac}
\affil[1]{Brookhaven National Laboratory, Upton, NY 11973-5000 USA}

\affil[*]{jzajac@bnl.gov}
\begin{abstract}
{Single photons are resources needed for developing quantum networks QN. They distribute
quantum information services across commercial optical fiber links and are key 
ingredient in developing quantum repeaters architectures. Currently, the most robust quantum light 
sources are Quantum Dots made of III-V materials. They emit highly indistinguishable photons on-demand 
and with high efficiency. Established devices work at near-infrared wavelengths (NIR) and further research 
is needed to develop devices working in telecommunication wavelengths O- and S-bands. 
In this contribution, we propose and model a broadband optical antenna working in O-band. 
It exhibits high extraction efficiencies with small Purcell enhancement around 2. 
We also examine far field emission from these structures, ensuring Gaussian mode profile is observed. }
\end{abstract}

\setboolean{displaycopyright}{false} 

\begin{document}
\maketitle
\section{Introduction}
Efficient light extraction from quantum emitters and increased light-matter interaction is 
crucial for developing robust optical links, which are fundamental building blocks of long 
distance quantum networks\cite{KimbleNature08,HansonWehnerScience18}. There had been extensive research on this subject 
dedicated to QDs working in the near-infra-red (NIR) spectral range. 
One of the approaches reported are solid-immersion lenses (SIL) or its derivatives micro-SIL and super-spheres. 
These SILs act as optical antennas by increasing coupling between emitter and light field, which 
relies on increasing the acceptance angle of radiation into an objective and thus numerical aperture (NA). 
In the past, high light extraction efficiencies of SILs and $\mu$SILs had been demonstrated for NIR 
III-V QDs\cite{MaZajacOptLett15, MaleinPRL16}, NV-centers in diamond\cite{HaddenRarityAPL10} and other material 
systems\cite{BekkerAPL23}.
In order to further improve light extraction efficiency more complex cavity structures had been used.
These include $\mu$SILs used as a top mirror and metal layers used as a bottom mirror\cite{MaZajacOptLett15}. 
Moreover, using designs with $\mu$SIL, as compared to macroscopic SIL, had 
been demonstrated in the past to suppress leaky modes. 
Recently, there had been reports on adapting different types of microcavity designs into telecom wavelengths 
exploiting 
Distributed Bragg Reflectors (DBRs) at 1.3$\mu$m\cite{BlokhinUsinovOptEx21,SrockaSekAIPAdv18}. 
%
%
\nolinenumbers
In this Letter, we have been investigating two designs; the first one consists of 
$\mu$SIL and Au bottom membrane with QD in the center of the SIL and the second one of super-$\mu$SIL and Au membrane.
Both of these designs are optimized for 1.3 $\mu$m.
%
\section{Design and Methodology}
Firstly, we consider InAs QDs in the center of hemisphere made of  
InAlGaAs quaternary alloy. 
Radius of the $\mu$SIL is defined as $R = r \cdot n/1350\,nm$ where $n$ is refractive index. 
Low temperature refractive index is estimated as n=3.2\cite{ZielinskaRudnorudzinskiOptExpress22}. 
Dimension of the semiconductor membrane is  $h = 311 $nm. At the bottom of the structure, we 
place Au layer $d=200$\,nm. This layer acts as both bottom mirror and contact for the Shottky diode. 
Au has negligible absorption at these wavelengths while we chose this thickness to limits leaky-modes. 
Second design is a super-sphere with characteristic dimension $a + r = (1 + \frac{1}{n})$, where $a$ is the 
distance from the top layer to the center of the super sphere, $r$ is the radius of the sphere, 
and $n$ is the refractive index. 
The simulation was carried out using Ansys Lumerical FDTD solver where we scan the radius for fixed wavelength ranges 
between 1200 and 1500 nm in order to find cavity-modes. We used perfectly matched layer 
(PML) boundary condition to truncate the FDTD simulation region, which has dimensions 10 x 10 x 4 $\mu$m 
\begin{figure}[h!]
\centering
\includegraphics[width=0.8\linewidth]{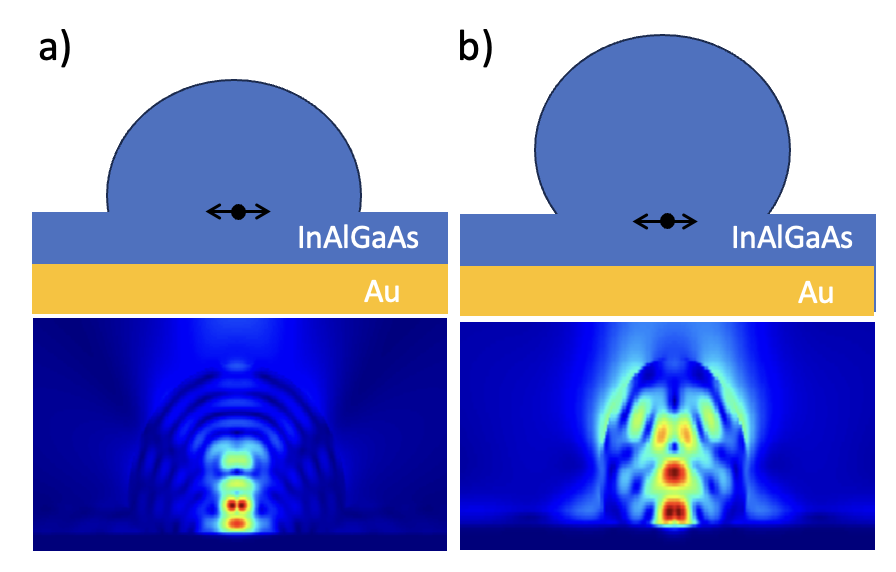}
\caption{In a) top row, layout of $\mu$SIL device with QD in the center and Au layer underneath. 
In the bottom row, field distribution within the $\mu$SIL structure.
In b) top row, the same but for super-$\mu$SIL. In the bottom row, field distribution within the super-$\mu$SIL.}
\label{device_layout}
\end{figure}
in the x, y, and z directions, respectively. We use power monitor in z-direction to obtain transmission power flux 
for the structure. 
We extract $\eta_{ex}$ - photon extraction efficiencies which are directly proportional to transmissions normalized 
to source powers, $F_p$ - Purcell enhancement and $\eta_{ph}$, where 
$\eta_{ph}$ = $\eta_{ex} \cdot F_p$. Moreover, we extract far-field profiles of electric fields.
Far field distributions were found by projecting data from a frequency-domain profile and power monitor to the far 
field using Lumerical script function farfield3d, which returns electric field intensity |E|$^2$ projected onto a 
hemisphere. For each mode, a 10x10 $\mu$m 2-dimensional z-normal monitor was placed above the SIL
and electric field data was collected. Finally, the relationship between transmission and the
numerical aperture (NA) of the monitor for several R values was found by integrating the far field projection over a 
cone for a range of half-angles between 0-$90^\circ$. 
The fraction of far field integrated for each angle was then multiplied by the Purcell factor to obtain $\eta_{ph}$ 
for this mode.
Finally, parameters had been collected into a table to compare performance of these two designs.
Device layout is shown in  Fig.\ref{hemi_all2}
\section{Results}
%
\begin{figure}[b!]
	\centering
	\includegraphics[width=1\linewidth]{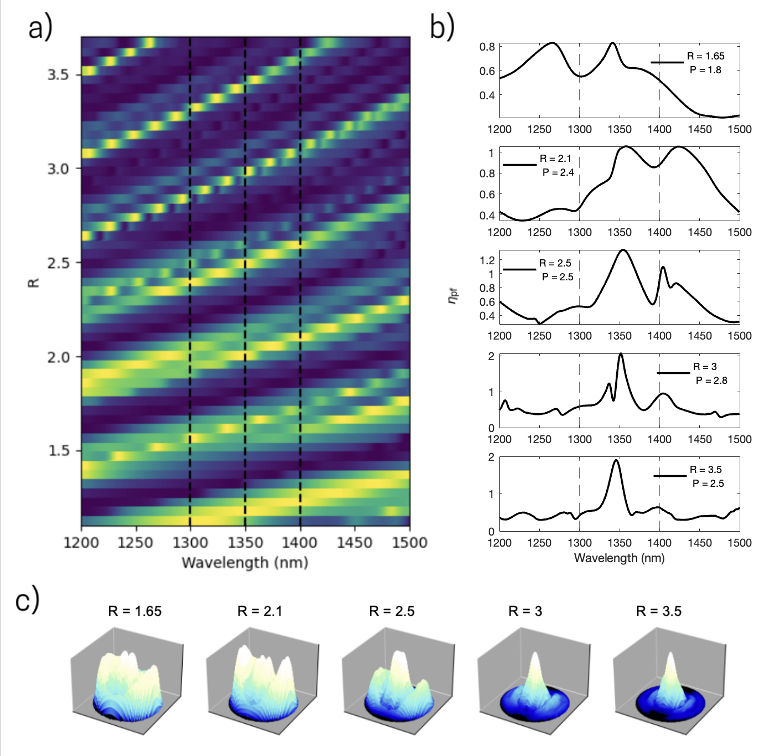}
	\caption{a) transmission maps vs radius (R) and wavelength\,(nm), b) Transmission cross-sections for fixed radius as a 
    function of wavelength, c) far-field mode profiles for hemisphere based design.}
	\label{hemi_all2}
\end{figure}
\begin{figure}[t!]
	\centering
	\includegraphics[width=0.8\linewidth]{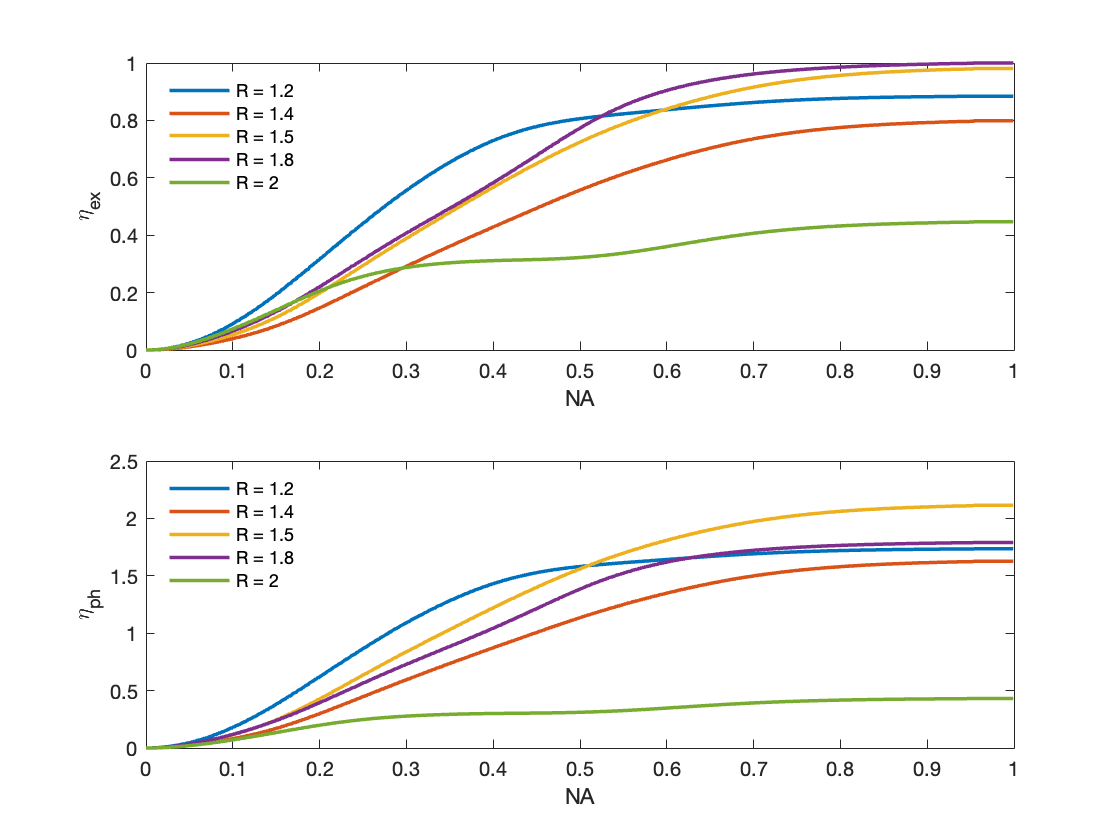}
    \caption{Top, transmission, $\eta_{ex}$, bottom, Purcell enhanced transmission for super-sphere based structure. 
    Purcell enhancement is mode dependent and between 1-2 for current design. For more details compare text.}
	\label{super_NA_map}
\end{figure}

In the Fig.\ref{hemi_all2} a), we are showing transmission maps as a function of radius and wavelength of the cavity for the 
case of a cavity wit hemisphere design. For these modes, we select a cross-section along the 1350\,nm central 
wavelength of interest and further plotting these cross=section in the Fig.\ref{hemi_all2} b). These data are showing 
broadband transmission with fitted FWHM vales between 15 and 45\,nm as given in Tab.\ref{table1}. Finally, far field profiles are 
given in Fig.\ref{hemi_all2} c) where the Gaussian modes identified for $R_H$ = 1.8  and $R_H$ = 3.7. 
%
%
\begin{figure}[t!]
	\centering
	\includegraphics[width=0.8\linewidth]{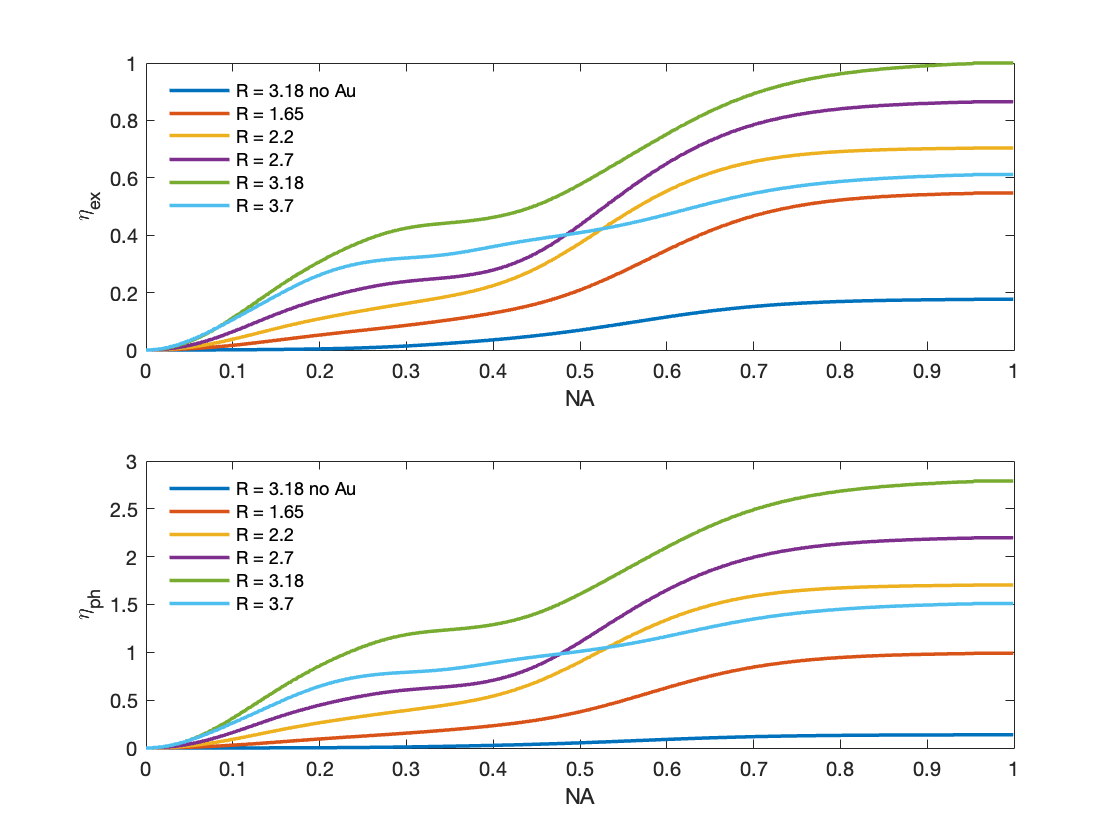}
    \caption{Top, transmission, $\eta_{ex}$, bottom, Purcell enhanced transmission for hemisphere based structure, $\eta_{ph}$. 
    For more details compare text.
    }
	\label{hemi_NA_map}
\end{figure}
In the Fig.\ref{hemi_NA_map} we are showing extraction efficiencies, $\eta_{ex}$, and Purcell enhanced extraction efficiency, 
$\eta_{ph}$. We identify two modes, for $R_H$ = 1.8 and $R_H$ =3.7, the earlier has higher $\eta_{ph}$. 
%
\begin{figure}[b!]
	\centering
	\includegraphics[width=1\linewidth]{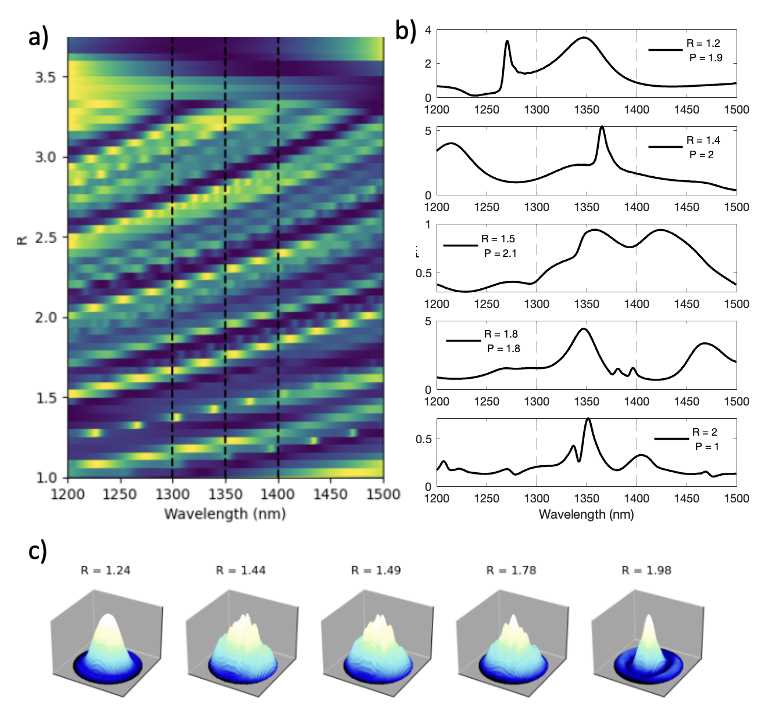}
	\caption{a) transmission maps vs radius (R) and wavelength\,(nm), b) Transmission cross-sections for fixed radius as a 
    function of wavelength, c) far-field mode profiles for hemisphere based design.}
	\label{super_all}
\end{figure}
In the Fig.\ref{hemi_all2} a), we show transmission maps as a function of radius and wavelength of the cavity for the 
case of a cavity wit hemisphere top mirror. For these modes, we preselect cross-section along the 1350\,nm central 
wavelength of interest and plot them in the Fig.\ref{hemi_all2} b). These data are showing 
broadband transmission with averaged FWHM as fitted with Gaussian fit function of around 30 nm as given in Tab.\ref{table1}. %
Finally, far field profiles are given in Fig.\ref{hemi_all2} c) with the most Gaussian modes identified for $R_H$ = 3  and $R_H$ = 3.5. 
In the Fig.\ref{hemi_NA_map} we are showing extraction efficiencies, $\eta_{ex}$ at the top, and Purcell enhanced extraction 
efficiencies at the bottom, $\eta_{ph}$. 
%
%
In Fig.\ref{super_all} we show in a) transmission maps, b) transmission cross-sections and c) far-field profiles 
for the $\mu$super-sphere design. While in Fig.\ref{super_NA_map} b), we plot transmission as function of NAs. 
As compared to hemisphere, modes for super-sphere have overall higher transmission at lower NA and this can be seen 
in Table.\ref{table1}. This result is consistent with literature \cite{ZwillerNJPhys04,SartisonPortalupiGiessenLAM21}. 
However, we do not see higher overall $\eta_{ex}$ for super-sphere design. This can be due to the fact that we use a 
bottom Au layer that offsets this effect. There is no trend observed when comparing $F_P$ for both designs.
For transmission maps in Fig.\ref{super_all}, cavity breaks for R values beyond 3.2.
This is explained by super-$\mu$sphere geometry. Namely, we defined $a + r$ having a fixed 
thickness of 311 nm. Once $r$ is varied beyond that, $a$ becomes zero and it causes collapse of the cavity. 
One can add another 311 nm for this case what should preserve cavity modes for higher R values calculations. 
%
%
%
\setlength{\tabcolsep}{4pt}
\begin{table}[t!]
\begin{tabular}[1\textwidth]{ccccccccccc}
\hline
\hline
&   &         &             &             &            &$\eta_{ex}$  &    &    &$\eta_{ph}$  &               
\\
\vspace{-0.05cm}
&R$_H$           & $\lambda_c$ & $\Delta \lambda$  & $F_P$  & 0.5 & 0.8 & 1 & 0.5 & 0.8 & 1      
\\
\hline 
& $1.65$  &    1339     &   44        & 1.81   &  0.2   &  0.5  & 0.54       & 0.37   &  0.94  & 1       
\\
& $2.2$   &    1355     &   15        & 2.42   &  0.37  & 0.7   & 0.7        & 0.9    &  1.7   & 1.7     
\\
& $2.7$   &    1355     &   33        & 2.54   &  0.43  & 0.84  & 0.86       & 1.1    &  2.1   & 2.2     
\\
&$3.18$  &    1350     &   20        & 2.79   &  0.58  & 0.96  & 1          & 1.6    &  2.7   & 2.8     
\\
&$3.7$   &    1345     &   15        & 2.47   &  0.4   & 0.59  & 0.6        & 1      &  1.45  & 1.5     
\\
\hline
&R$_S$           & $\lambda_c$ & $\Delta \lambda$  & $F_P$  & 0.5 & 0.8 & 1 & 0.5 & 0.8 & 1      
\\
& 1.24  & 1338        &   43        & 1.9  & 0.8    &  0.87   & 0.88       & 1.6   &  1.7  & 1.74          
\\
 & $1.44$  & 1367        &   14        & 2    & 0.55   &  0.77   & 0.8        & 1.1   &  1.6  & 1.62       
\\
& $1.49$  & 1356        &   40        & 2.1  & 0.72   & 0.96    & 1          & 1.56  & 2     & 2.1        
\\
& $1.78$  & 1348        &   25        & 1.8  & 0.77   &  0.98   & 1          & 1.38  &  1.8  & 1.9        
\\
& $1.98$  & 1350        &   22        & 1    & 0.32   &  0.4    & 0.44       & 0.3   &  0.4   & 0.4       
\\
\hline
\hline
\end{tabular}
\caption{ Comparison of performance for hemispherical (top part of table) and super-spherical (bottom part of table) 
$\mu$SILs cavities for different modes marked according to R and for NA values of 0.5, 0.8 and 1.}
    \label{table1}
\end{table}

In conclusion, we have presented two designs for O-band QDs cavity structure ensuring efficient extraction of photons. 
Super-sphere design ensures higher extraction efficiencies for smaller NA what can be useful for experiments with small 
NA of collection optics. However, due to the bottom Au layer, there is no overall higher collection efficiencies by using super-SIL 
as for the case without the membrane.
For both designs, far field modeling show that there are good quality Gaussian modes, namely, for 
$R_H = 3 and 3.5$ and $R_S = 1.98$. For the later, it is also noted that this $R_S$ mode have low extraction efficiency and 
better choice good be to use $R_S = 1.24$ even though Gaussian shape is compromised.
We have considered here case of Quantum Dots, but this can be easily extended to other single emitters. 
\\ JMZ acknowledge support from Department of Energy, US through Carrier Grant (2023-2028).
\section{Disclosures} ABC: 123 Corporation (I,E,P), DEF: 456 Corporation (R,S). GHI: 789 Corporation (C).
\\The authors declare no conflicts of interest.

\section{Data availability} Data underlying the results presented in this paper are not publicly available at this time but may be obtained from the authors upon reasonable request.

\bibliography{Refs}

\newpage
\textbf{References}
\begin{enumerate}
    \item H. J. Kimble, Nature 453 (2008) \textit{The Quantum Internet}
    \item S. Wehner, D. Elkouss, and R.Hanson, Science 362 (2018) \textit{Quantum internet: A vision for the road ahead}
    \item Y. Ma, G. Ballesteros, J. Zajac, et al., Opt. Lett. 40, 2373 (2015) \textit{Highly directional emission from a quantum emitter embedded in a hemispherical cavity}
    \item R. N. E. Malein, T. S. Santana, J. M. Zajac, et al., Phys. Rev. Lett. 116,257401 (2016) \textit{Screening Nuclear Field Fluctuations in Quantum Dots for Indistinguishable Photon Generation}
    \item J. P. Hadden, J. P. Harrison, A. C. Stanley-Clarke, et al., Appl. Phys. Lett. 97, 24 (2010) \textit{Strongly enhanced photon collection from diamond defect centers under microfabricated integrated solid immersion lenses}
    \item C. Bekker, M. J. Arshad, P. Cilibrizzi, et al., Appl. Phys. 122 (2023) \textit{Scalable fabrication of hemispherical solid immersion lenses in silicon carbide through grayscale hard-mask lithography}
    \item S. A. Blokhin, M. A. Bobrov, N. A. Maleev, et al., Opt. Express 29, 5 (2021) \textit{Design optimization for bright electrically-driven quantum dot single-photon sources emitting in telecom O-band}
    \item N. Srocka, A. Musiał, P.-I. Schneider, et al., AIP Avdances 8, 085205 (2018) \textit{Enhanced photon-extraction efficiency from InGaAs/GaAs quantum dots in deterministic photonic structures at 1.3 $\mu m$ fabricated by in-situ electron-beam lithography }
    \item A. Zielinska, A. Musiał, P. Wyborski, et al., Opt. Express 30, 12 (2022) \textit{Temperature dependence of refractive indices of Al0.9Ga0.1As and In0.53Al0.1Ga0.37As in the telecommunication spectral range}
    \item V. Zwiller, T. Aichele, and O. Benson, New J. Phys. 6, 96 (2004) \textit{Quantum optics with single quantum dot devices}
    \item  M. Sartison, K. Weber, S. Thiele, et al., Light. Adv. Manuf. 2, 6 (2021) \textit{3D printed micro-optics for quantum technology: Optimised coupling of single quantum dot emission intoa single-mode fibre}

\end{enumerate}
\end{document}